\documentstyle[epsf]{l-aa}
\newcommand{\qua}{RX\,J0947.0+4721}
\newcommand{\wea}{RX\,J0947.1+4721}
\newcommand{\Hf}{\mbox{$\rm{H}_0\,=\,50\frac{km}{s\,Mpc}$}}
\newcommand{\qn}{\mbox{$\rm{q}_0$}}
\newcommand{\fie}{HS 47.5/22}
\newcommand{\oa}{\mbox{$\rm \phi_{oa}$}}
\newcommand{\NH}{\mbox{$\rm{N}_{\rm{H}}$}}
\newcommand{\lan}{\langle}
\newcommand{\ran}{\rangle}
\newcommand{\ergss}{\mbox{$\,\rm{ergs}\,\rm{s}^{-1}$}}
\newcommand{\cps}{\mbox{$\,\rm{cts}\,\rm{s}^{-1}$}}
\newcommand{\kms}{\mbox{$\,\rm{km}\,\rm{s}^{-1}$}}
\newcommand{\ergscms}{\mbox{$\,\rm{ergs}\,\rm{cm}^{-2}\,\rm{s}^{-1}$}}
\newcommand{\ergscmshz}
           {\mbox{$\,\rm{ergs}\,\rm{cm}^{-2}\,\rm{s}^{-1}\,\rm{Hz}^{-1}$}}
\begin{document}
      \thesaurus{03 (11.17.3;  11.17.4(\qua ); 11.17.4(\wea ); 13.25.2 )}
   \title{\qua\ -- an extremely soft Narrow Line QSO 
      \thanks{Based partly on observations from the German-Spanish Astronomical 
      Center, Calar Alto, operated by the Max-Planck-Institut f\"ur Astronomie, 
      Heidelberg, jointly with the Spanish National Commission for Astronomy}
          }
   \subtitle{}
   \author{K. Molthagen\thanks{\emph{present address:} MPI f\"ur extraterresrtr. Physik, 
              D 85740 Garching, Germany}
           \and N. Bade \and H.J. Wendker } 
   \offprints{K. Molthagen, kmolt@mpe.mpg.de}
   \institute{Hamburger Sternwarte, Gojenbergsweg 112, D -- 21029 Hamburg, 
              Germany      }
   \date{Received ; accepted }
   \maketitle
   \begin{abstract}
    \qua\ is a newly discovered example of a narrow--line Seyfert 1 type object
    with a quasar level luminosity ($\rm M_B\,=\,-24.8$). It was detected in
    the ROSAT All--Sky Survey (RASS) and later in 17 pointed observations of
    the ROSAT Position sensitive Proportional Counter (PSPC). The optical
    spectrum contains strong FeII emission.
    We apply several spectral models to the X--ray data. A single power law fit
    yields a photon index $\rm\Gamma\,=\,-4.2$. Single component thermal
    models underestimate the flux above 1\,keV. A good fit is obtained with
    a blackbody plus power law model ($\Gamma\,=\,1.7$ fixed) modified by
    Galactic absorption. The best fit temperature is 
    $\rm T_{bb}\,=\,108\pm8\,eV$, and the flux ratio in the ROSAT band is
    $\rm f_{bb}/f_{pl}\,=\,6$. 
    Large count rate variations are visible with a maximum amplitude of a
    factor $>\,17.5$. The fastest change is a drop of a factor 2.7 in 25 hours.
    No spectral variations were observed.
    The maximum and minimum luminosities 
    in the two component model are $\rm 1.0\cdot10^{46}\,\ergss$ and 
    $<\,6\cdot10^{44}\,\ergss$, respectively. We discuss possible explanations 
    for the extreme softness and the observed variability. 
   \end{abstract}
\section{Introduction}
Studies of the soft X--ray excess were done preferentially with low redshift 
AGN because this spectral feature will be shifted out of the 
observed energy windows of the ROSAT, EXOSAT and {\it Einstein} satellites
for higher redshifts.
The existence of the soft X--ray excess was first detected for 
the $\rm z\,=\,0.038$ Seyfert 1 galaxy E1615+061 (Pravdo et al. 1981) and
Mrk 841, another 
Seyfert galaxy with $\rm z\,=\,0.036$ (Arnaud et al. \cite{arnaud}). Among 
the 58 AGN used for a study of the ultraviolet to soft X--ray bump (Walter 
\& Fink, \cite{walterfink}) the QSO with the highest redshift is 3C\,263 with 
$\rm z\,=\,0.652$, and only one of the 31 narrow line Seyfert 1 galaxies 
(NLS1) investigated by Boller et al. (1996), has z higher than that. 
This object, E1346$+$266 (see also Puchnarewicz et al. 1994), and PHL\,1092 
(Forster \& Halpern 1996) are the most distant narrow line type 1 QSOs 
reported 
so far. It is therefore of great interest to examine the properties of 
further narrow line QSOs with high redshifts and steep soft X--ray spectra. 
The situation is similar for the study of the time behaviour of AGN. For a 
statistically unambiguous detection of a short--term variation a minimum of 
source counts is necessary. Therefore such studies preferred bright nearby AGN.

\qua\ was first detected in the RASS\footnote{{\bf R}OSAT {\bf A}ll {\bf S}ky 
{\bf S}urvey (Voges 1992)} (Ba\-de et al. 1995). It appeared to be a very soft 
source, but the photon statistics were insufficient to determine the
spectral shape. Besides that, \qua\ is the brightest and steepest source in
a medium deep ROSAT survey in the field \fie\ (Molthagen 1996). The field is 
part of the Hamburg Quasar Survey (Hagen et al. 1995), an objective prism 
based survey of the northern hemisphere. Due to its brightness, \qua\ can be
found in 16 of the 48 pointings which form the survey. Strong variations, 
already suggested in the RASS data, were clearly seen in the pointed data.

\qua\ shows striking similarity in optical and X--ray behaviour 
with NLS1 objects such as IRAS\,13224$-$3809 (Boller et al. \cite{boller}). 
Both \qua\ and IRAS\,13224$-$3809 have exceptionally steep soft X--ray 
spectra, narrow Balmer lines, and strong 
optical Fe\,II emission. \qua\ is also an IRAS source (Moshir et al. 1990). Its 
optical and X--ray luminosities, computed with \Hf\ and $\qn\,=\,0$, clearly 
denote it as quasar. The low Galactic absorption in the direction of this 
QSO enables high--quality spectral analysis of this quasar--lumi\-nosity NLS1.
\section{The Data}
\subsection{X--ray observations \label{Xdatq38}}
\qua\ was detected during an extensive search for AGN in the RASS by Bade 
et al. (\cite{bade2}). They found a very soft source with a count rate 
of $0.10\pm 0.02$\cps\ in the total ROSAT band (0.1 -- 2.4 keV).

Besides its presence in the RASS and the medium deep survey pointings, 
RX\,J0947.0+4721 is located in the outer part of a pointing on the Abell 
cluster A851. All observations were made with the PSPC detector 
(Pfef\-fer\-mann et al. 1986) on board the ROSAT satellite (Tr\"umper 1983), 
between April 1991 and October 1993. Each pointing was split into at least 
two observational intervals (OBIs) separated by intervals ranging from a few 
hours to more than a year. The OBIs are therefore treated separately for 
timing analysis. Table \ref{pdat} lists date and length of the OBIs as well 
as number of counts, off--axis angle \oa, count rates and hardness ratios 
of \qua\ found therein.

The entrance window of the PSPC is covered by a thin foil supported by aluminum
struts (Pfeffermann et al. 1986). Depending on its position in the field of
view (FOV), these struts can shadow a source. ROSAT is operated in
the so--called `wobble mode' to lessen the imprints of the struts. As a
consequence, sources close to the detector edge can be moved (partly) out of
the FOV. \qua\ is obscured by the struts or moved partly out of the FOV in
several OBIs. These are marked (-) in Table \ref{pdat}.

In pointing 700165, \qua\ was detected only in the second OBI, but not in the
first, although it is unobscured in both. We determined the vignetting 
corrected number of counts between 0.1 and 1.0\,keV in the source area of 
the first OBI, $\rm N_{tot}\,=\,100.6\,cts$, and use the count rate 
$\rm cr\,=\,3\sqrt{N_{tot}}/t_{exp}\,=\,0.029\,\cps$ as an upper limit in the
following.
\begin{table*} 
\caption[]
     {Observation log of \qua. Count rates are given in the 0.1 -- 1.0 keV band.
     Partially shadowed OBIs are marked (-).} 
\label{pdat}
\begin{tabular}{l|rrrrll}
\hline
 \multicolumn{1}{c|}{Pointing} & \multicolumn{1}{c}{date} & 
 \multicolumn{1}{c}{$\rm t_{exp}$} &\multicolumn{1}{c}{$\rm N_{ct}$} 
 &\multicolumn{1}{c}{$\phi_{\rm oa}$} &
 \multicolumn{1}{c}{$\rm cr_{\rm broad}$} & \multicolumn{1}{c}{HR} \\
 &  &\multicolumn{1}{c}{[\,s\,]} & & \multicolumn{1}{c}{[\,$'$\,]}
 & {\small[$\rm 10^{-2}\frac{cts}{s}$] } &  \\
\hline
700166-1    & 91 04 14&1826 &  626.2&12.6  &$34.3\pm1.5$    &$-0.80\pm0.05$ \\ 
700171-1    & 91 04 14&1732 &  418.4&37.1  &$24.2\pm1.6$    &$-0.76\pm0.08$ \\ 
700171-2    & 91 04 18&591  &  299.5&37.1  &$50.7\pm4.0$    &$-0.80\pm0.10$ \\ 
700166-2    & 91 04 19&377  &   71.0&12.6  &$18.8\pm2.7$    &$-0.88\pm0.19$ \\ 
700170-1 (-)& 91 04 20&1386 &  163.1&52.1  &$11.8\pm2.0$    &$-0.74\pm0.21$ \\ 
700167-1 (-)& 91 04 21&571  &   64.3&22.9  &$11.3\pm1.7$    &$-0.82\pm0.20$ \\ 
700172-1    & 91 04 21&1097 &  204.6&8.7   &$18.7\pm1.4$    &$-0.79\pm0.10$ \\ 
700180-1 (-)& 91 04 21&1125 &  160.1&45.4  &$12.4\pm1.6$    &$-0.77\pm0.17$ \\ 
700173-1 (-)& 91 04 23&1521 &  158.7&25.7  &$10.3\pm1.0$    &$-0.75\pm0.12$ \\ 
700182-1 (-)& 91 05 10&837  &   93.7&23.5  &$11.2\pm1.4$    &$-0.75\pm0.15$ \\ 
700182-2 (-)& 91 05 10&2027 &  210.2&23.5  &$10.4\pm0.9$    &$-0.78\pm0.11$ \\ 
700165-1    & 91 11 15&1047 &$<\,30$&41.0  &$\ \,2.9^a$     &  \\ 
700181-1 (-)& 91 11 15&2263 &$<\,36$&43.6  &$\ \,1.6^a$     &  \\ 
800102-1 (-)& 91 11 17&1298 &   63.6&47.9  &$\ \,4.9\pm1.4$ &$-0.71\pm0.35$ \\ 
700165-2    & 91 11 17&943  &   86.0&41.0  &$\ \,9.1\pm1.6$ &$-0.81\pm0.23$ \\ 
800102-2 (-)& 91 11 18&2400 &  249.3&47.9  &$10.3\pm1.4$    &$-0.85\pm0.18$ \\ 
800102-3 (-)& 91 11 18&1904 &  147.7&47.9  &$\ \,7.6\pm1.3$ &$-0.87\pm0.23$ \\ 
800102-4 (-)& 91 11 19&1401 &  180.5&47.9  &$12.9\pm1.8$    &$-0.67\pm0.17$ \\ 
800102-5 (-)& 91 11 19&2253 &   89.1&47.9  &$\ \,3.8\pm1.4$ &$-0.76\pm0.46$ \\ 
800102-6 (-)& 91 11 19&2114 &  191.0&47.9  &$\ \,9.0\pm1.4$ &$-0.76\pm0.19$ \\ 
800102-7 (-)& 91 11 19&1634 &   68.6&47.9  &$\ \,4.2\pm1.3$ &$-1^b$ \\ 
800102-8 (-)& 91 11 20&1198 &   72.8&47.9  &$\ \,6.1\pm1.7$ &$-0.71\pm0.35$ \\ 
700181-2 (-)& 91 11 20&673  &   76.3&43.6  &$11.3\pm2.3$    &$-0.96\pm0.28$ \\ 
700457-1    & 92 05 09&1665 &  642.6&39.6  &$38.6\pm2.0$    &$-0.70\pm0.06$ \\ 
700454-1 (-)& 92 05 09&1585 &  266.7&24.1  &$16.3\pm1.2$    &$-0.71\pm0.09$ \\ 
700451-1    & 92 05 10&1695 &  408.8&43.7  &$24.1\pm1.8$    &$-0.73\pm0.09$ \\ 
700453-1    & 92 05 10&1768 &  428.4&34.1  &$24.2\pm1.5$    &$-0.78\pm0.08$ \\ 
700453-2    & 92 05 18&2887 &  662.8&34.1  &$23.0\pm1.2$    &$-0.76\pm0.06$ \\ 
700454-2 (-)& 92 05 19&1590 &  327.8&24.1  &$20.6\pm1.4$    &$-0.82\pm0.09$ \\ 
700457-2 (-)& 92 05 19&1607 &  397.9&39.6  &$24.8\pm1.7$    &$-0.82\pm0.09$ \\ 
700451-2 (-)& 92 05 19&1726 &  453.9&43.7  &$26.3\pm1.8$    &$-0.84\pm0.09$ \\ 
700173-2 (-)& 92 11 14&566  &   97.8&25.7  &$16.9\pm2.1$    &$-0.76\pm0.15$ \\ 
700180-2    & 92 11 19&696  &   73.3&45.4  &$10.5\pm2.0$    &$-0.93\pm0.25$ \\ 
700168-1 (-)& 92 11 19&716  &   75.5&30.5  &$10.5\pm1.6$    &$-0.89\pm0.20$ \\ 
700180-3    & 92 11 20&600  &  107.1&45.4  &$16.2\pm2.3$    &$-0.66\pm0.17$ \\ 
700167-2 (-)& 92 11 20&700  &   49.7&22.9  &$\ \,7.1\pm1.2$ &$-0.74\pm0.21$ \\ 
700170-2 (-)& 92 11 21&1032 &  100.7&52.1  &$\ \,9.8\pm2.1$ &$-0.94\pm0.30$ \\ 
700169-1 (-)& 93 04 24&1380 &  122.5&53.4  &$\ \,8.9\pm2.0$ &$-0.94\pm0.31$ \\ 
700169-2 (-)& 93 04 25&1018 &   81.6&53.4  &$\ \,8.0\pm2.2$ &$-0.57\pm0.32$ \\ 
700168-2    & 93 04 25&791  &  237.3&30.5  &$30.0\pm2.4$    &$-0.69\pm0.10$ \\ 
700168-3    & 93 04 25&619  &  167.0&30.5  &$27.0\pm2.6$    &$-0.81\pm0.12$ \\ 
700167-3 (-)& 93 10 20&1451 &  237.2&22.9  &$16.3\pm1.2$    &$-0.86\pm0.10$ \\ 
\hline
\multicolumn{4}{l}{ \footnotesize $^a$: upper limit for non--detection} &
\multicolumn{3}{l}{ \footnotesize $^b$: no source counts above 0.4\,keV}\\
\end{tabular}
\end{table*}

\setlength{\unitlength}{1mm}
\begin{figure}[hbtp]
 \begin{picture}(88,87)
  \put(5,5){\epsfbox{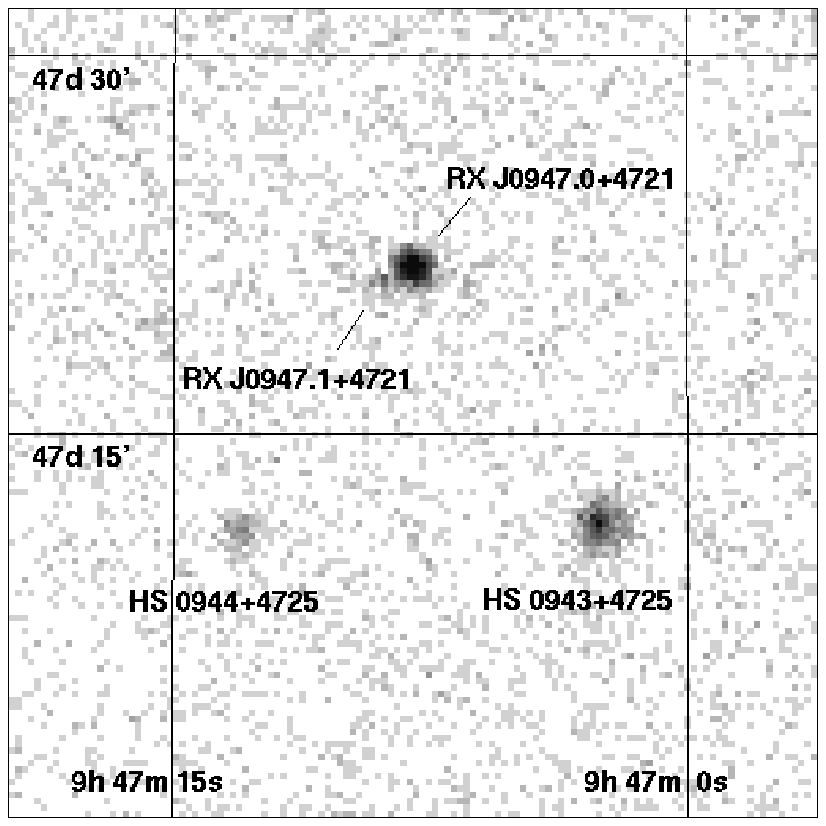}}
  \put(15,-1){\epsfbox{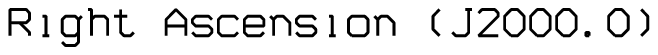}}
  \put(0,22){\epsfbox{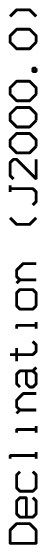}}
 \end{picture}
 \caption[]
   {Merged image in the total ROSAT band (0.1 -- 2.4 keV) showing the four 
    AGN \qua, \wea, HS0943$+$4725 ($\rm z\,=\,0.233$, Bade et al. 1995; Engels 
    et al. 1997) and HS0944$+$4725 ($\rm z\,=\,0.703$, Engels et al. 1997). 
    Only pointings with \qua\ located in the inner $15'$ of the 
    PSPC FOV are merged.} 
  \label{q38i}
\end{figure}
In those OBIs where \qua\ is located closest to the detector centre 
(\oa\ $<\,15'$), a fainter and harder source was found at a distance of 
$1'\, 22''$ from \qua\ (Fig. \ref{q38i}). The count rate of this 
neighbouring source, \wea, is $\rm 0.015\pm 0.002\,cts\,s^{-1}$ in the 
total ROSAT band. With the adaptive hardness ratio method developed by 
Schartel et al. (1996), assuming a power law spectrum 
($f_{\rm x}\,\sim\,\nu^{-\Gamma}$) with absorption fixed at the Galactic 
value, we found a 
$\Gamma\,=\,2.41^{+0.25}_{-0.28}$ for \wea , and a flux of 
$\rm f_x\,=\,(\,1.62\pm0.24\,)\cdot10^{-13}\,\ergscms$ in the total ROSAT 
band, which is about a factor 10 lower than the overall average value of \qua\ 
(see next section). The source might even be slightly fainter and harder than 
the values computed here, because the two sources are not fully separated.

In all other pointings the two sources are blended, due to the large point 
spread function (PSF) at larger off--axis angles, so that any extraction 
circle around \qua\ includes at least parts of the weaker source's photons. 
The contamination 
will be more significant at the higher energies where \qua\ is faintest. The 
analysis was therefore largely restricted to the 0.1 -- 1.0 keV band 
(0.15 -- 1.54 keV in the quasar's rest frame) which is called 'broad' in the 
following. 
\subsection{Optical observations}
\begin{figure}[htbp]
 \epsfbox{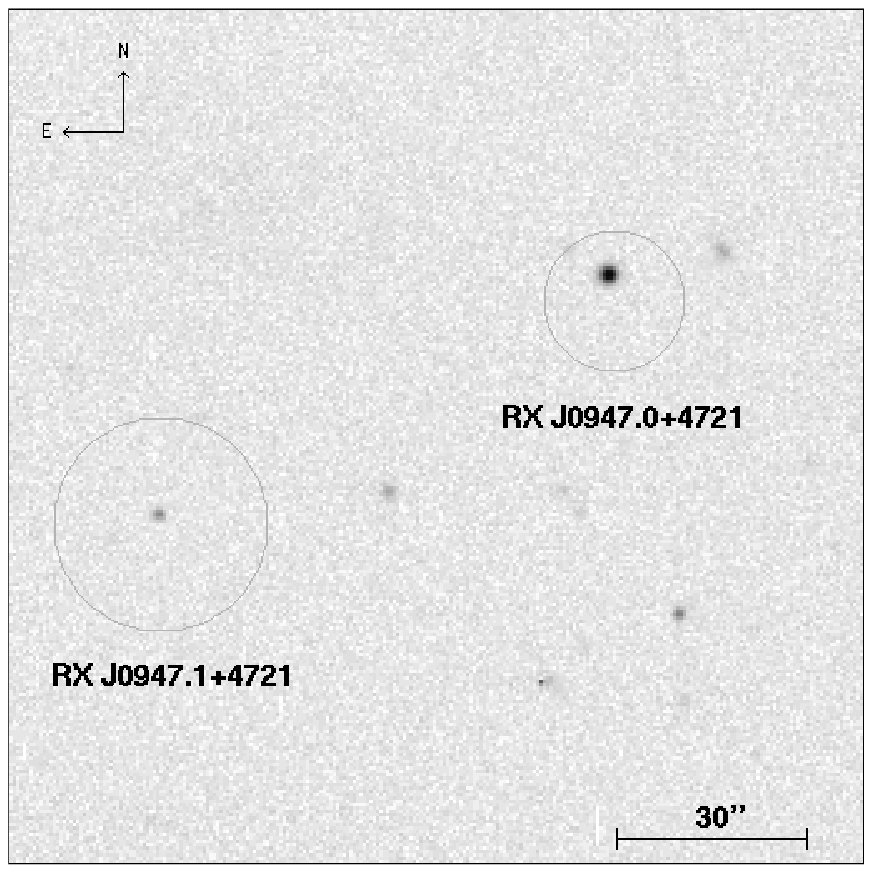}
 \caption[]
   {60s blue CCD (Tektronix, $\rm 0.53''/pix$) image including \qua\ and
    \wea . The circles represent the X--ray positional errors. } 
   \label{qdir}
\end{figure}

Figure \ref{qdir} shows a blue CCD image including both \qua\ and \wea . It 
was taken with the 3.5m telescope $+$ focal reducer at Calar Alto in March 
1994. The B--magnitudes for the two objects are $18\fm11\pm0\fm04$ and 
$21\fm1\pm 0\fm4$, respectively. In both cases, optical and X--ray positions 
agree within $6''$, and no other optical sources are present in the X--ray 
error circles, so that the identifications can be considered as unambiguous.

The identification spectrum of \qua\ was taken in February 1992 
(Bade et al. \cite{bade2}). 
It shows Mg\,II and narrow Balmer emission lines with a redshift of 
$\rm z\,=\,0.541$ and strong bumps due to the Fe\,II multiplets around 
$\rm 4570\,\AA$ and $\rm 5250\,\AA$.

\begin{figure}[htbp]
 \epsfbox{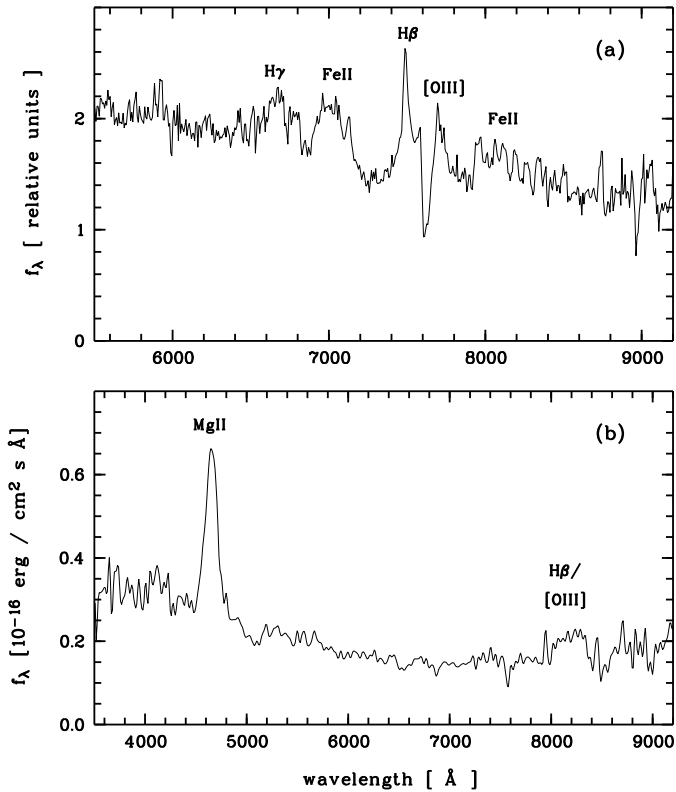}
 \caption[]
   {Spectra of (a) RX\,J0947.0+4721 and (b) \wea. $\rm f_{\lambda}$
    is plotted against observed wavelength. 
    Both spectra were taken in March 1994 with the 3.5m telescope at Calar
    Alto, equipped with focal reducer and two different grisms. } 
   \label{ca94}
\end{figure}

An additional spectrum with $\rm35\,\AA$ resolution taken at red wavelengths 
has the range from H$\gamma$ to the FeII$\lambda 5250$ bump in the centre 
(Fig. \ref{ca94}a). Due to oncoming clouds during the exposure, no absolute 
flux calibration is possible, only relative values are given. H$\beta$ is of 
similar width and strength as in the first spectrum: $\rm EW\,=\,23.7\,\AA$ 
and $\rm FWHM\,=\,1370\pm170\,\kms$. (The instrumental profile has been 
subtracted. The quasar did not fill the slit, and so the FWHM might be 
slightly underestimated.) The rest frame equivalent widths for H$\beta$ and 
FeII$\lambda 4570$ are given in Table \ref{odat}. The value for 
FeII$\lambda 4570$ might be underestimated due to the presence of the sky 
absorption line at $\,6880\,\rm \AA$ at the blue side of the bump. Their ratio 
is FeII$\rm \lambda4570/H\beta\ \geq\ 1.75$, placing \qua\ among the strong 
Fe\,II emitters, where strong means FeII$\rm \lambda4570/H\beta\ \geq\ 0.5$ 
(Joly \cite{jol2}). 

\wea\ was observed with a low dispersion grism ($\rm 905\,\AA/mm$) to obtain 
high signal to noise in a short time. The spectrum is shown in Fig. 
\ref{ca94}b. It contains one strong emission line at $\rm 4670\,\AA$ which 
we identified with Mg\,II, leading to a redshift $\rm z\,=\,0.67$. 
The broad bump between $7990\,\rm \AA$ and $\rm 8416\,\AA$ would then contain 
H$\beta$ and the [OIII] lines. No further emission lines can be seen in the
spectrum, and since no strong lines are expected between MgII and H$\beta$, we
are confident that our identification is correct.
\begin{table}[htbp]
\caption{Parameters of \qua. \NH: own measurements (Molthagen et al. 
    \cite{pacat}); equivalent widths: in the QSO's rest frame.}
\label{odat}
\begin{tabular}{c|c}
\hline 
 & \qua \\
\hline 
  R.A. (J2000.0) & $\rm 9^h\ 47^m\ 4\fs2$  \\ 
  DEC. (J2000.0) & $47^{\circ}\ 21'\ 42''$ \\
  $\rm m_b$ & $18\fm11\,\pm\,0\fm04$  \\ z & 0.541 \\
  $\rm M_b$ & $-24.8$ \\
  N$_{\rm H,gal}$$^a$ & $\rm 1.07\cdot 10^{20}\,cm^{-2}$ \\
  EW$_{\rm H\beta}$ & 23.7\,$\rm \AA$ \\  
  EW$_{\rm FeII\lambda 4570}$ & 41.4 $\rm \AA$  \\  
\hline
\multicolumn{2}{l}{$^a$: error $\simeq\,20\%$}\\
\end{tabular} 
\end{table}
\subsection{Radio observations}
\qua\ was detected as a radio source with the Dominion Radio Astrophysical 
Observatory synthesis radio 
telescope\footnote{details at \,{\tt http://www.drao.nrc.ca/web/homepage.shtml}}
in Penticton, Canada. At 1420 MHz we found a source with a flux density
$\rm f_{1420}\,=\,4.8\pm1.2\,mJy$, $24''$ apart from the optical position. This 
is within the expected uncertainty of the radio position. Measurements were 
also made at 408 MHz, but due to confusion with a nearby source only 
an upper limit of $\rm f_{408}\,<\,20\,mJy$ was obtained. 

With these two values, an upper limit for the radio spectral index 
($f_{\nu}\,\sim\,\nu^{-\alpha_{\rm R}}$), $\alpha_{\rm R}\,<\,1.14$, can be
estimated. It is thus not impossible that \qua\ has a flat radio spectrum, i.e. 
$\rm\alpha_R\,<\,0.5$.
A limit for the radio--loudness $R_{\rm L}$ can be estimated from the upper
limit of $\rm\alpha_{R}$. Wilkes \& Elvis (1987) consider a quasar to be 
radio--loud if $R_{\rm L}\,=\,\log\left(f_{\rm 5GHz}/f_{\rm B}\right)\,>\,1$.
With $\alpha_{\rm R}\,<\,1.14$ and $\rm m_B$ from Table \ref{odat}, we get 
$\rm f_{\rm 5\,GHz}\,>\,1.14\cdot10^{-26}\ergscmshz$ and 
$\rm f_{\rm B}\,=\,2.559\cdot10^{-27}\ergscmshz$, leading to 
$\rm R_L\,>\,0.65$. This is quite close to the dividing value, and depending 
on the true spectral shape, R$_{\rm L}$ can be larger than one. 

The quasar is also found in the 1.4\,GHz NRAO VLA Sky Survey (Condon et al.
\cite{nvss}). Its flux there is $\rm f_{1.4GHz}\,=\,3.6\pm0.4\,mJy$, 
consistent with our DRAO value.
\section{Results \label{Resu}}
\subsection{Temporal Behaviour \label{Tbehav}}
\begin{figure*}[ht]
 \epsffile{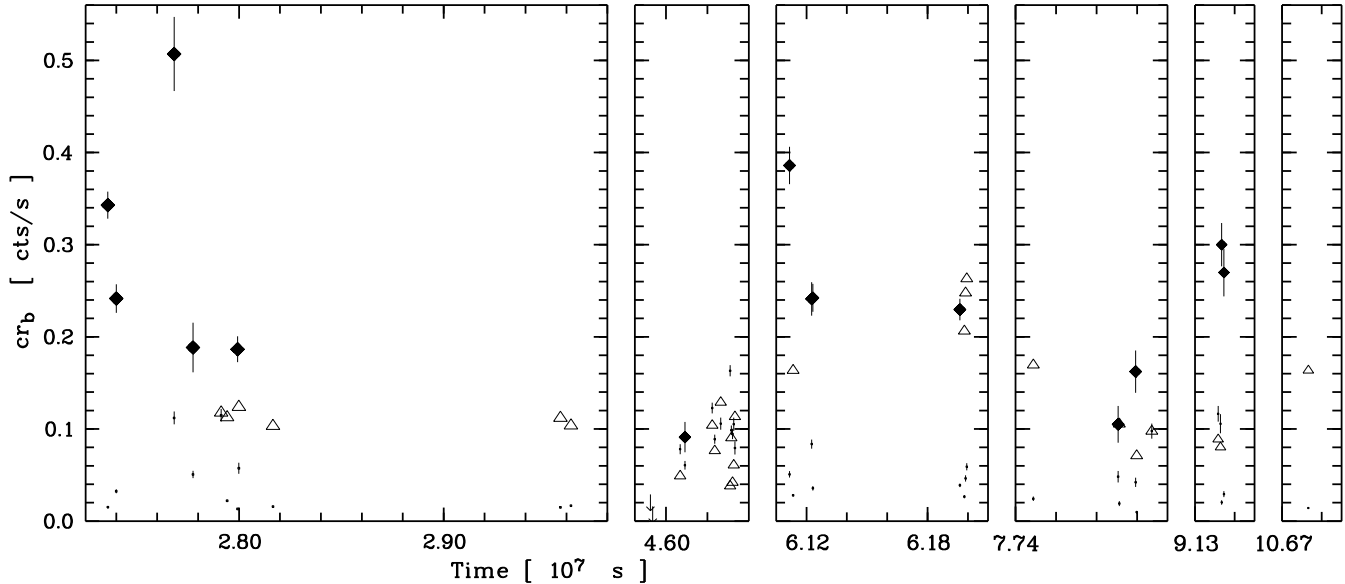}
 \caption[Light curve of \qua ]
   {Light curve of \qua. Broad band count rate is plotted against time in
    spacecraft seconds. Each entry represents one OBI. Filled diamonds: good
    OBIs, open triangles: obscured OBIs, arrows: upper limits for 
    non--detections, dots: background. } 
 \label{q38lk}
\end{figure*}

The light curve of \qua, shown in Fig. \ref{q38lk}, contains data from 
all OBIs. The largest 
amplitude of variability between two unobscured detections is a factor of 
$5.7\pm0.5$. However, this is not the maximum amplitude observed. 
Since there is no technical reason for the non--detection in OBI 700165-1, 
a genuine change of a factor $\rm>\,17.5$ must have occured.  

Besides these large variations, the count rate has been seen to drop by a 
factor of 2.7 in 25 hours as well as a factor of 1.6 in 30 hours. There are 
also less significant indications of more rapid variability. Unfortunately, 
some of the relevant OBIs are obscured. The count rates derived from these 
data can be underestimated, so that the variations are of questionable 
significance.

It has to be checked  whether the changes are pure intensity variations or 
spectral variations as well. Since the OBIs contain generally too few hard 
photons to perform a meaningful spectral fit, the hardness ratio is used 
instead, which is defined as $$ H\!R\ =\ \frac{H\,-\,S}{B}\ \ \ ,$$ with B, 
H and S being the counts in the broad, hard (0.4 -- 1.0 keV) and soft 
(0.1 -- 0.4 keV) band, respectively. The error in HR is then 
$$\sigma_{HR}\ =\ B^{-1}\cdot\sqrt{\sigma_{\rm H}^2\,+\,\sigma_{\rm S}^2\,
                  +\,\left(H\!R\,\sigma_{\rm B}\right)^2}\ \ .$$

There could be some dependence of the hardness ratio on the position in the 
FOV: at large off--axis angles, parts of the soft PSF might fall outside the 
detector and the source appears harder. The available data did not show any 
correlation between HR and \oa, regardless of whether the OBIs are partially 
shadowed or not. Separate unweighted averages for obscured and unobscured 
OBIs yield $\rm \lan HR_{obs}\ran\,=\,-0.79\pm0.09$ and 
$\rm \lan HR_{unobs}\ran\,=\,-0.78\pm0.07$, respectively, the overall 
average is $\rm \lan HR\ran\,=\,-0.79\pm0.08$. 

\begin{figure}[htbp]
 \epsffile{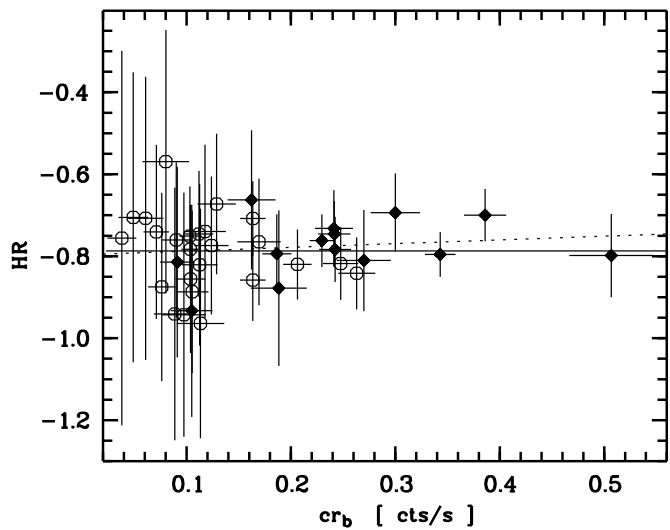}
 \caption[HR versus cr.]
   {Hardness ratio HR versus broad band count rate. OBIs without valid HR
    are left out. Open symbols: obscured
    data; solid line: average HR; dotted line: straight line fit to all data} 
 \label{hrcr}
\end{figure}
Figure \ref{hrcr} shows HR vs. count rate. Three OBIs have been left out, the
two upper limits and 800102-7, where no hard source counts were detected.
No trend of HR with $\rm cr_b$ can be seen. A $\chi^2$ fit to all data (39
OBIs) gives $H\!R\,=\,(-0.80\pm0.05)\,+\,(0.09\pm0.19)\cdot cr$ with
$\chi^2\,=\,8.84$ and a likelihood that this value of $\chi^2$ will be
exceeded by chance of $\rm q\,=\,1.000$. Fits to the good (14) and
obscured (25) OBIs separately give 
$H\!R\,=\,(-0.78\pm0.08)\,+\,(-0.08\pm0.28)\cdot cr$, $\chi^2\,=\,3.56$, 
$\rm q\,=\,0.991$ and
$H\!R\,=\,(-0.75\pm0.05)\,+\,(-0.31\pm0.47)\cdot cr$, $\chi^2\,=\,4.24$, 
$\rm q\,=\,0.999$, respectively. 
The $\chi^2$ values for constant HR are 9.50, 4.14 and 4.67 for all, good, and
obscured OBIs. In all cases, the slopes are compatible with zero, and the
differences in $\chi^2$ are too small to favour the straight line fit.
The observed changes are therefore considered to be pure intensity variations.
\subsection{Spectral Behaviour \label{Sbehav}}
Since no changes in HR could be detected, all data were merged into one 
spectrum ($8971\pm150\rm\,cts$) before conducting spectral fits. The spectrum 
was binned with roughly the same relative error 
$\sigma_{\rm cr} / cr\,\simeq\,6\%$ in each bin below 1.0\,keV, yielding 14 
bins. Above 1.0\,keV, only $277\pm38$ source photons are detected; these 
are filled into one single bin with a relative error of 14\%.
Merging, binning, and fitting was done using the EXSAS software package
(Zimmermann et al. 1993); no additional errors were added. 
\begin{table*}[tbp]
\caption {This Table lists the results of fitting several models to
   \qua. Fluxes are observed values, kT is given in the quasar's rest frame.} 
\label{xspec}
\begin{tabular}{l|lcrlrcc}
\hline
& \multicolumn{1}{c}{N$_{\rm H}$} & $f_{\rm x}$ 
        & \multicolumn{1}{c}{kT} & $\Gamma$  
        & \multicolumn{1}{c}{L} & $\chi^2/d.o.f.$ &d.o.f.\\
& [$\rm 10^{20}\,cm^{-2}$] & [$10^{-12}\frac{\rm erg}{\rm s\,cm^2}$] 
        & \multicolumn{1}{c}{[ eV ]} & &[$\rm 10^{45}\,\frac{erg}{s}$] & &\\
\hline
power law &$1.07^a$ & $1.95\pm0.05$ & & $3.3\pm0.1$ & 4.0\hspace*{3mm} & 4.66 
          & 12 \\
{\hspace*{6mm}$''$} & $2.32\pm0.45$ & $6.04\pm2.11$ & &$4.2\pm0.9$
          & 12.3\hspace*{3mm} &1.24 & 11 \\
thermal bremsstrahlung& $1.07^a$ & $1.54\pm0.04$ & $308\pm17$ 
          & &3.1\hspace*{3mm} &1.69 & 12 \\
{\hspace*{6mm}$''$} & $1.51\pm0.36$ & $2.05\pm0.44$ & $265\pm25$ 
          & &4.2\hspace*{3mm} &1.19 & 11 \\
blackbody & 1.07$^a$ & $1.48\pm0.04$ & $116\pm\ \,3$ 
          & &3.0\hspace*{3mm} & 1.32 & 12 \\
{\hspace*{6mm}$''$} & $0.74\pm0.30$ & $1.25\pm0.12$ & $121\pm18$ 
          & &2.5\hspace*{3mm} & 1.20 & 11 \\
\hline
bb + power law$^{b,c}$ & 1.07$^a$ & $1.32\pm0.05,\ 0.23\pm0.06$ 
        & $103\pm\ \,8$ 
        & $1.7^d$ &3.1\hspace*{3mm} & 1.08 & 12 \\
bb + power law$^{b,c,e}$ & 1.07$^a$ & $2.15\pm0.10,\ 0.35\pm0.10$ 
        & $108\pm\ \,8$ 
        & $1.7^d$ &5.1\hspace*{3mm} & 1.40 & 12 \\
\hline
\multicolumn{2}{l}{\footnotesize $^a$: N$_{\rm H}$ f\,ixed at the Galactic value}&
\multicolumn{3}{l}{\footnotesize $^b$: f\,itted range is 0.1 -- 2.4 keV} &
\multicolumn{3}{l}{\footnotesize $^c$: $\rm f_{bb},\ f_{pl}$ given separately} \\
\multicolumn{2}{l}{\footnotesize $^d$: f\,ixed at the 'canonical' value }&
\multicolumn{6}{l}{\footnotesize $^e$: only unobscured OBIs used}\\
\end{tabular}
\end{table*}

\begin{figure}[htbp]
 \epsfbox{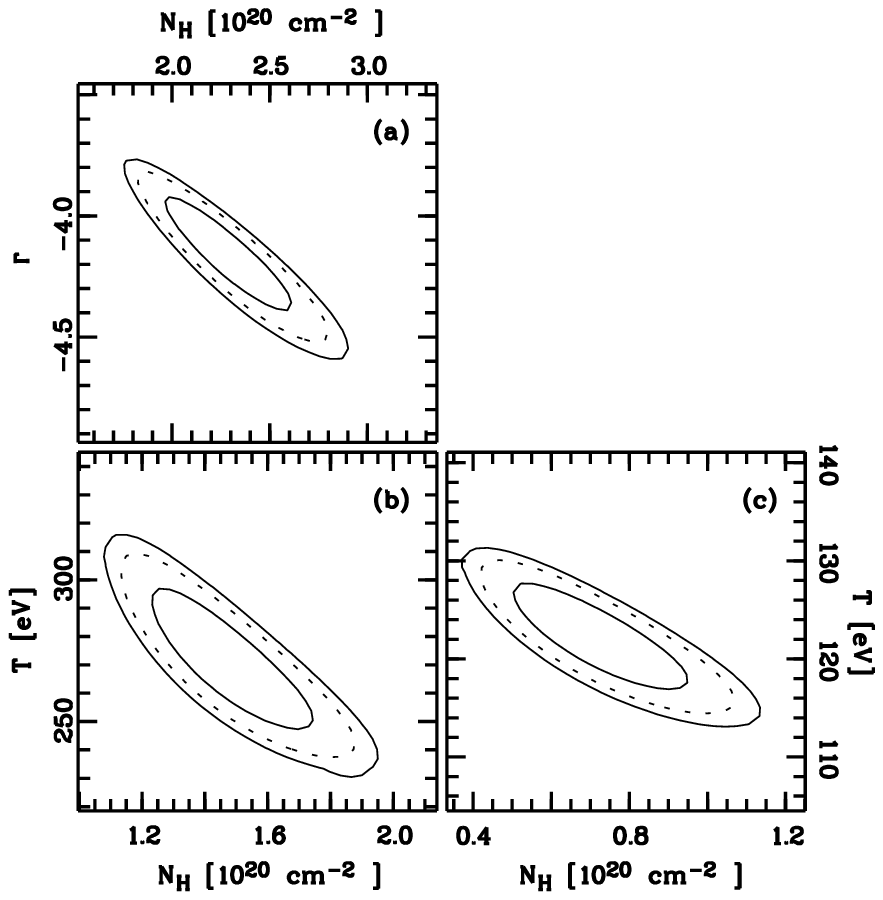}
 \caption[]
   {$\chi^2$ contour plots ($68\%$, $90\%$ and $95\%$ confidence level) for the 
    single component fits presented in Table \ref{xspec}. The indicated spectral
    parameter is plotted against \NH (in $\rm 10^{20}\,cm^{-2}$). \\
    {\bf a:} single power law; {\bf b:} thermal bremsstrahlung; {\bf c:} 
    blackbody.}
 \label{chc}
\end{figure}

First, several single--component models were applied to the 0.1 -- 1.0 keV range.
Table \ref{xspec} lists the results of the fits, and Fig. \ref{chc} shows 
the error ellipses.
A single power law with cold absorption gives an unacceptable fit 
when N$_{\rm H}$ is fixed at the Galactic value. With N$_{\rm H}$ free,  
the value obtained for the column density is more than twice the Galactic 
value. Thermal bremsstrahlung with fixed Galactic absorption yields a 
rather poor fit with $\rm\chi^2/d.o.f.\,=\,1.7$ and $\rm q\,=\,0.06$. 
Treating \NH\ as a free parameter improves the fit, but again results in 
\NH\,$>\,\rm N_{H,gal}$. An absorbed blackbody model gives acceptable fits 
for N$_{\rm H}$ both fixed and free. The Galactic N$_{\rm H}$ is higher than 
the best fit value, but still within the $90\%$ confidence interval. 

When the high energy tail (1.0 -- 2.4 keV) is included, both thermal models 
underestimate the flux above 1 keV significantly. A single power law fits the 
data, but again the fitted N$_{\rm H}$ is more than twice the Galactic value. 

\begin{figure}[htbp]
    \epsfbox{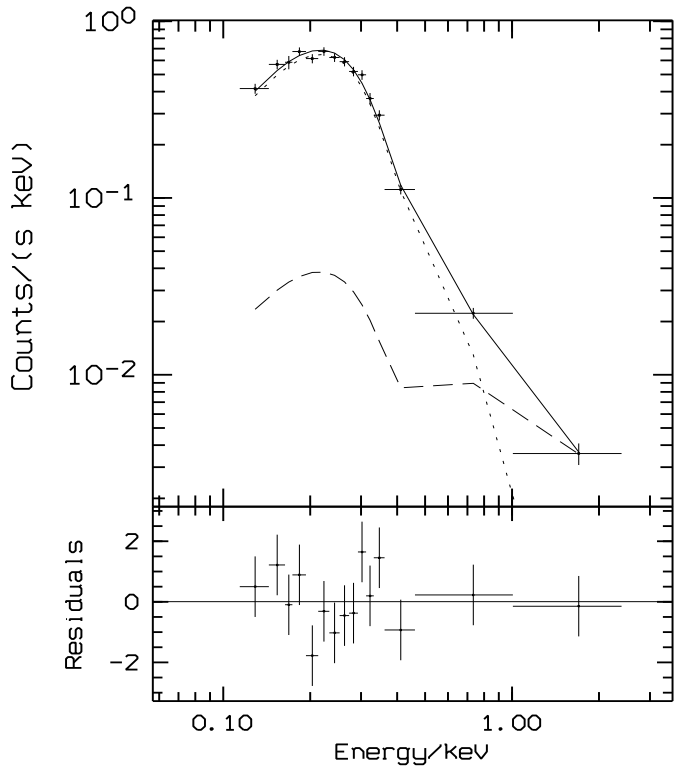}
   \caption[]
     {Fit of an absorbed blackbody plus power law model to the average
      spectrum of \qua. The dotted and dashed lines mark the blackbody
      and power law components, respectively.} 
   \label{b+p}
\end{figure}

We then tried a two--component fit of the total ROSAT band with Galactic 
absorption, the soft component modeled by a blackbody and the hard by a power 
law. It might be thought that the hard component is mainly a result of 
contamination by \wea\ which is at least partly inside the extraction 
area of \qua. To check this, we fixed power law index and flux at the values 
estimated for that source. The fit gives unacceptable results by 
underestimating the high energy tail of the spectrum, thus 
hinting at a noticeable hard component of \qua\ itself.
We decided to fix the power law index at the 'canonical' value 
$\Gamma\,=\,1.7$ and fit only the two normalizations and the blackbody 
temperature because the statistics above 1\,keV are too poor 
to fit $\Gamma$ as well. Only $3\%$ of the source photons are detected above
1\,keV, and if these are filled into more than one spectral bin, the 
relative error per bin is above 25\%, too high to give reasonable constraints
on $\Gamma$. 
The fit result is presented in Fig. \ref{b+p}, and the parameters are 
listed in Table \ref{xspec}. 

With this model, the average flux is
$\rm f_x\,=\,(1.55\pm0.08)\cdot 10^{-12}\,erg\,s^{-1}\,cm^{-2}$
if all data are averaged, and 
$\rm f_x\,=\,(2.49\pm0.14)\cdot 10^{-12}\,erg\,s^{-1}\,cm^{-2}$
if only the unobscured OBIs are used. As was the 
case for the hardness ratios, no significant change in the spectral parameters 
can be seen when only the unobscured OBIs are used for the fits. 
The contribution of \wea\ to the overall averaged 
flux will be about $10\%$ at worst (section \ref{Xdatq38}).

With H$_0$ and q$_0$ as before, the fitted flux corresponds to a good (overall) 
average luminosity L$_{\rm x}\,=\,5.1\cdot10^{45}\ergss$ ($3.1\cdot10^{45}\ergss$)
in the QSO's rest frame (0.15 -- 3.70 keV). 
The detected maximum and minimum luminosities, in this model, are 
$\rm L_{max}\,=\,1.0\cdot 10^{46}\,\ergscms$ and 
$\rm L_{min}\,=\,1.8\cdot 10^{45}\,\ergscms$, respectively, and the upper limit
corresponds to $\rm L_x\,<\,6\cdot10^{44}\,\ergss$. We applied no 
K--correction to avoid including parts of the spectrum which we did not 
observe, and to avoid a shift of the band onto the Rayleigh--Jeans part of the 
blackbody component of the spectrum. The values for 
f$_{\rm x}$ and L$_{\rm x}$ have to be taken with caution, because they are 
strongly model dependent. 

\begin{figure}[htbp]
 \epsfbox{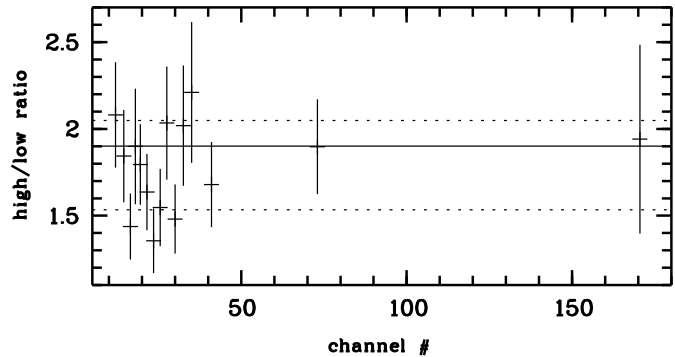}
 \caption[]
   {Ratio of `high count rate' to `low count rate' spectra (good data).
    Solid line: average value, dotted lines: average $\pm$ standard
    deviation.}
   \label{dhl}
\end{figure}

So far, no indication for spectral variability could be found. As a further
test, we created `high count rate' and `low count rate' spectra which were 
binned as described before, and then divided (fig. \ref{dhl}). The separating 
count rate, $\rm 0.24\,\cps$, was chosen to give roughly the same number of 
counts in both spectra. Spectral changes should be visible as a deviation of 
the ratio from a constant.
No such deviation can be seen. The data are consistent with a constant value 
($\rm\chi^2\,=\,16.6$ 13\,d.o.f.; $\rm q\,=\,0.987$). If all data are used, 
or if a different binning is applied, the results are similar. Separate 
fits to high and low spectra give spectral parameters in good 
$_{\rm x}\,=\,5.1\cdot10^{45}\ergss$agreement with the ones in Table 
\ref{xspec}. Again, the conclusion is that no significant spectral changes 
are seen.
\section{Discussion \label{Disk}}
\qua\ is denoted as a QSO (rather than a Seyfert) by its optical luminosity of
$\rm M_B\,=\,-24.8$. This is supported by the overall averaged X--ray 
luminosity of $3.1\cdot10^{45}\,\ergss$; Seyferts do not normally have 
$\rm L_x$ above several $10^{44}\,\ergss$.
Otherwise, \qua\ is very similar to NLS1 such as IRAS\,$13224-3809$ 
(Boller et al. 1993). The photon index and FWHM of $\rm H\beta$
place it well inside the range obtained by the NLS1 sample in Boller et al. 
(1996; their fig. 8). The ratio FeII$\lambda 4570$$/\rm H\beta\,\geq\,1.75$ is 
comparable with other values found for these objects. X--ray variability as 
observed is a common feature in NLS1 (Boller et al. 1996). As for
IRAS\,13224$-$3809, IZw1 and other NLS1, \qua\ has a high far infrared flux. 
It has been detected by IRAS at 60$\,\mu$ with a flux of $\rm 245\,mJy$ 
(Moshir et al. 1990), corresponding to an IR--luminosity of 
$\rm\nu F_{60\mu}\,=\,2.5\cdot 10^{46}\,\ergss$.  

\begin{table}
 \caption[]{Parameters of the narrow line QSOs.}
 \label{nlqso}
 \begin{tabular}{l|ccc}
  \hline
 & \qua & E1346+266$^a$ & PHL\,1092$^b$ \\
  \hline
 $\Gamma$ & $4.2\pm0.9$ & $4.1\pm0.2$ & 
  \begin{minipage}[t]{13mm}{$4.17^{+1.41}_{-1.10}$\vspace*{1mm}}
  \end{minipage}\\
 $\rm L_x^c\ [\frac{erg}{s}]$ & $0.7\ldots 4\cdot 10^{46}$ & 
  $\rm 2.4\cdot 10^{46}$ 
  &\begin{minipage}[t]{13mm}{$1.9\cdot 10^{46}$\vspace*{2mm}}\end{minipage} \\
  $\rm M_B$ & $-24.8$ & $-25.5$ & $-25.2$ \\
  FWHM$_{\rm H\beta}$ & $\rm 1300\pm170\,\frac{km}{s}$ & 
     $\rm 1840\,\frac{km}{s}$ & $\rm 1800\,\frac{km}{s}$ \\
  $\rm EW_{H\beta}\ [\AA] $ & $23.7$ & 31.1 & 14 \\
  $\rm Fe\,II/H\beta$ & $\geq\,1.75$ & 0.98 & 5.3 \\
  \hline
  \multicolumn{2}{l}{\footnotesize $^a$: Boller et al. 1996} &
     \multicolumn{2}{l}{\footnotesize $^b$: Forster \& Halpern 1996} \\
  \multicolumn{4}{l}{$^c$: \begin{minipage}[t]{80mm}{The values of 
      $L_{\rm x}$ are highly model dependent. Here, a power law with
      Galactic \NH\ (no K--correction) was applied, together with $\qn\,=\,0$,
      \Hf.}\end{minipage} }
 \end{tabular}
\end{table} 
A comparison of \qua\ with the narrow line quasars
shows that their X--ray parameters are quite similar: the luminosities in the 
ROSAT band are of the same order of magnitude, and the photon indices are 
practically identical. This latter observation may be taken as an indication 
that the steepness extends towards fairly high rest frame energies. However, 
the indication must not be overstressed. The value of $\Gamma$ for 
E1346$+$266 is not well determined; Puchnarewicz et al. (1994) find smaller
values ($\rm \Gamma\,=\,3.34\ldots 3.71$) than Boller et al. (1996). 

A better description of the observed spectrum can be achieved by thermal 
models, however, both the blackbody and thermal bremsstrahlung models 
underestimate the flux above 1\,keV. 
The analysis in section \ref{Xdatq38} has shown that the hard component of the 
spectrum cannot be totally explained with the neighbouring harder source 
which is too weak. It must therefore be intrinsic to RX\,J0947.0+4721.

A good fit is obtained with a blackbody modeling the soft excess plus a power 
law as hard component, modified by Galactic absorption. For an accretion 
disk model this means that the X--rays originate from a small region with 
little variations in temperature. 

This interpretation assumes a thermal origin for the X--ray emission in the 
inner parts of the accretion disk. There is, however, an indication against 
this assumption. No change of the hardness ratio with increasing count rate 
is observed, which means that the temperature of the emitting region remains 
constant over nearly one order of magnitude in luminosity. Over this 
luminosity range temperature variations should be easily detectable from the 
correlation $L \propto T^4$. 

Model calculations for a spectrum with fixed power law component, and all
flux changes attributed to a temperature change in the blackbody component, 
show that only small deviations from the average flux $\rm f_{ave}$ 
(table \ref{xspec}, last row) would be compatible with the data. 
\begin{figure}[htbp]
    \epsfbox{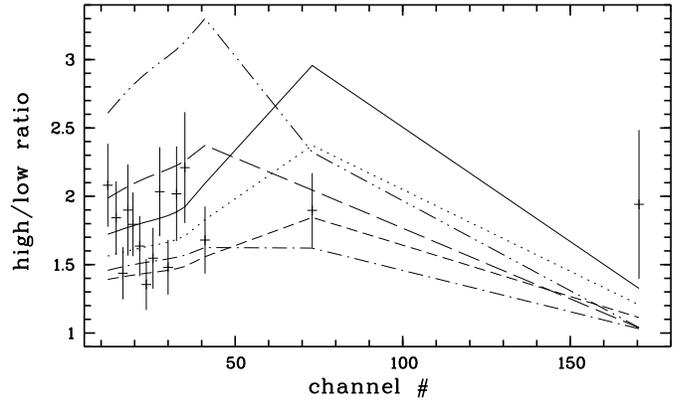}
   \caption[]
     {Ratio of high to low count rate spectra versus channel number. Crosses: 
      data; solid line: $\rm f/f_{ave}\,=\,2$, dotted line: 
      $\rm f/f_{ave}\,=\,1.75$, short dashed line: $\rm f/f_{ave}\,=\,1.5$, 
      dash-dotted line: $\rm f/f_{ave}\,=\,2/3$, long dashed line: 
      $\rm f/f_{ave}\,=\,0.5$, dash-dot-dotted line: $\rm f/f_{ave}\,=\,0.4$
      } 
   \label{spmod}
\end{figure}
This is illustrated in Fig. \ref{spmod}, which shows the ratio of high to 
low count rate spectra for the data (crosses with error bars) and six 
simulations (lines) with ratios $\rm f/f_{ave}$ between 0.4 and 2. The curves
for ratios 2, 1.75, and 0.4, show significant deviations from the observed
ratio, while the others agree with the data below channel 100 (roughly
1\,keV). Only changes less than a factor 1.75 -- 2 can be caused by a change
in T. The large changes observed (factors 5.7, 2.7) must be caused by another
process.

A luminosity change in dependency of the radius (pulsations with constant 
temperature, $L \propto R^2$) is difficult to model for an accreting black 
hole because the X--ray emitting region should have always the same distance 
(in Schwarzschild radii) from the central engine. As a consequence, models 
involving reprocessing of harder X--ray into softer photons 
could be a solution.

In Guilbert \& Rees (\cite{guilbert}), the central engine produces a 
non--thermal spectrum of hard X--rays and $\gamma$--rays of sufficient energy 
to produce electron -- positron pairs which in turn can produce and maintain 
a secondary $\rm e^+ - e^-$ plasma with optical depth $>\,1$. The large 
optical depth is responsible for the existence of a cool component in the 
gas close to the central engine. Incident radiation is reflected and 
reprocessed by cold material.
More recent works (e.g. Maraschi \& Haardt 1997, Haardt et al. 1997) have
shown that the disk corona may have only electrons instead of $\rm e^{+}e^{-}$
pairs, and that it is not necessarily optically thick. The fact that the iron
$\rm K\alpha$ line is not Comptonized to invisibility supports
$\rm\tau\,<\,1$ for the corona. However, although these models can explain the
shape of the soft X--ray spectrum, the lack of spectral variations may cause
problems. In Haardt et al. (1997), noticeable spectral changes are predicted
for the ROSAT band, none of which are observed for \qua. 

Models including warm absorbers may be another possible explanation of NLS1 
phenomena. A detailed description of such a model applied to Mrk\,766 can 
be found in Leighly et al. (\cite{m766}). Unfortunately, Mrk\,766 is more 
a Sy\,1.5 than a NLS1 (Osterbrock \& Pogge 1985), and it may be 
inappropriate to generalize the results for this special object to the 
whole NLS1 class. The application of such a model to \qua\ would conflict 
with the lack of spectral variability, anyway. 
Changes of ionisation parameter and/or column density of the warm absorber  
would cause changes in spectral shape.  
Spectral changes are expected even if the variations arise via changes in 
the central source itself since photoionisation contributes to the ionisation 
structure of the absorber. 
The lack of spectral changes for \qua\ is a strong argument against warm
absorber models.
 
Further model constraints can in principle be derived from timing analysis.
The shortest variation significantly detected for \qua\ corresponds to a 
decrease $\rm \Delta L\,=\,6\cdot10^{45}\,\ergss$ (bb + pl model) in 
$\rm 9.2\cdot 10^5\,s$. Following Fabian \& Rees (1979), this value implies an 
efficiency $\eta\,\geq\,5\cdot10^{-43}\,\Delta L\,/\Delta t\,=\,0.003$ -- far 
below the limit of 0.057 for accretion onto a Schwarzschild black hole. If the 
single power law model is used instead, we have 
$\rm \Delta L\,=\,2.4\cdot 10^{46}\,\ergss$, and subsequently 
$\rm \eta\,\geq \,0.013$, still more than a factor 4 below the limit.

Blazar--like activity, known to produce rapid X--ray variations, is usually not 
applied to NLS1 galaxies because 
NLS1 objects do not show properties which are typical for jet activities, as
there are flat spectrum radio emission, strong polarisation, and nearly
featureless multifrequency spectra. \qua\ may be different in that 
the possibility of a flat radio spectrum (i.e. $\rm\alpha_R\,\leq\,0.5$, 
which would make it also radio loud) can presently not be ruled out. 

If the short time variations are confirmed, a Schwarzschild black hole will no 
longer be a proper model. The Schwarzschild limit is passed if 
large amplitude variations ($\rm \Delta L\,=\,2\cdot10^{45}\,\ergss$)
in 16000\,s are detected.  These would suggest either the presence of a 
Kerr black hole or relativistic X--ray beaming effects. 

The large amplitude variation of a factor $>\,17.5$ hints at models like a 
filamentary or spot--like emission region (IRAS\,13224$-$3809; Otani et al. 
1996) or tidal disruption of stars (IC\,3599; Brandt et al. 1995, Grupe et 
al. 1995) which have been suggested to explain giant amplitude variations in 
these ultrasoft NLS1 galaxies. 
\section{Summary}
This paper presents optical and soft X--ray observations of a QSO discovered 
within the RASS and present in 16 pointings from a medium deep ROSAT survey 
in the field \fie\ (Molthagen et al., \cite{pacat}). 

The object is similar to narrow--line Seyfert 1 galaxies, but both its 
optical and X--ray luminosity are about an order of magnitude higher than most 
of the objects in Boller et al. (1996), thus clearly extending the object class 
into the quasar domain.
 
It shows an extremely soft X--ray spectrum. The usual description with a power 
law and low energy absorption gives only very poor fits. Single component 
thermal models underestimate the flux above 1\,keV. A two component model with 
the soft excess modeled by a blackbody, and a power law as hard component, 
gives a better fit. Strong and -- at the high luminosity -- remarkably fast 
variations can be seen in the light curve of \qua.

Warm absorbers could be an 
explanation of NLS1 phenomena. Applied to \qua, such models conflict with the 
lack of observed spectral variability. However, the faintness of \qua\ at 
harder X--rays may have prohibited the detection of changes close to 1\,keV.

Models including reprocessing might be another explanation for the 
X--ray emission of \qua, but again, the lack of spectral variations may be a
problem.

At present, the hard X--ray data are not sufficient to distinguish between
the above models, or others like a filamentary or spot--like emission region,
or the tidal disruption of stars. A thorough investigation of intensity {\em
and} spectral changes would be highly  desirable.
\begin{acknowledgements} 
   The authors wish to thank W.N. Brandt for many useful comments. 
   KM acknowledges support by BMFT (DARA FKZ 50 OR 9308) and 
   NB by the Deutsche Forschungsgemeinschaft under Re 353/22-1 to 4 and DARA
   50\,OR\,960116.
\end{acknowledgements} 

\end{document}